# Hidden topological signatures in a van der Waals ferromagnet revealed by spin-resolved circular-dichroic photoemission

Jay R. Paudel[1,2], Federico Mazzola[3,4], Lidia A. Lapinski[1,5], Barun Ghosh[6,7], Valeria Rosa Rocha[8], Abigail M. Derrico[1,8], Nikola Maksimovic[8], Iulia Cojocariu[9], Daniel Baranowski[9], Christoph Klewe[10], Vitaliy Feyer[9], Ivana Vobornik[11], Claus M. Schneider[9], Giancarlo Panaccione[11], Arun Bansil[6,12], James G. Analytis[8], and Alexander X. Gray[1,*]

[1] *Department of Physics, Temple University, Philadelphia, Pennsylvania 19122, USA*

[2] *Chemical Sciences Division, Lawrence Berkeley National Laboratory, Berkeley, California 94720, USA*

[3] *Istituto di Struttura della Materia, Consiglio Nazionale delle Ricerche (CNR-ISM), 34149 Trieste, Italy*

[4] *Department of Physics and Astronomy "Galileo Galilei", University of Padova, 35131 Padova, Italy*

[5] *Department of Physics, Stanford University, Stanford, California 94305, USA*

[6] *Department of Physics, Northeastern University, Boston, Massachusetts 02115, USA*

[7] *Department of Condensed Matter and Materials Physics, S. N. Bose National Centre for Basic Sciences, Kolkata, India*

[8] *Department of Physics, University of California, Berkeley, California 94720, USA*

[9] *Peter Grünberg Institut (PGI-6), Forschungszentrum Jülich GmbH, 52428 Jülich, Germany*

[10] *Advanced Light Source, Lawrence Berkeley National Laboratory, Berkeley, California 94720, USA*

[11] *Istituto Officina dei Materiali, Consiglio Nazionale delle Ricerche (CNR-IOM), 34149 Trieste, Italy*

[12] *Quantum Materials and Sensing Institute, Northeastern University, Burlington, Massachusetts 01803, USA*

[*] *E-mail: axgray@temple.edu*

Layered ferromagnets with strongly correlated electrons provide a platform for exploring the interplay among magnetism, spin-orbit coupling, and nontrivial band topology. $Ni_{1/4}TaSe_2$ is a correlated van der Waals ferromagnet with a transition temperature of 58 K that lies close in composition to a superconducting phase that emerges upon Ni depletion. Here, we combine spin- and angle-resolved photoemission spectroscopy, circular-dichroism measurements, and density functional theory calculations to investigate its spin-resolved electronic structure. We observe spin-polarized states crossing the Fermi level and signatures of broken time-reversal symmetry in the ferromagnetic state. Spin-resolved circular-dichroism measurements further reveal a helicity-dependent spin response consistent with a finite spin Berry curvature and suggestive of local topological character. Together, these results establish an experimental connection between ferromagnetism, spin-polarized electronic states, and topological signatures in $Ni_{1/4}TaSe_2$, providing a foundation for investigating how these properties may relate to the emergence of superconductivity upon Ni depletion.

Correlated itinerant ferromagnetism is among the most intriguing electronic phases in solids, where the interplay between electron correlations and magnetic order can drive electronic instabilities toward unconventional states such as triplet superconductivity [1,2], non-Fermi-liquid behavior [3], and topological flat-band states [4,5]. Understanding how electronic correlations and magnetism interact with spin-orbit coupling and band topology is therefore an important route toward uncovering new quantum phases, particularly in low-dimensional materials, where reduced dimensionality can enhance interaction effects and make spin-orbit-driven phenomena especially prominent.

Van der Waals ferromagnets provide an attractive platform for exploring this interplay in relatively simple, well-defined systems. Their weak interlayer bonding enables exfoliation and integration into spintronic devices [6,7]. Of particular interest is the interaction between spin-orbit coupling (SOC) and magnetic order, which can give rise to chiral phenomena [8] and anomalous Hall responses [9]. The intrinsic anomalous Hall response is governed by the Berry curvature of the electronic states [9,10], providing a direct link between magnetism and topological properties.

Among low-dimensional van der Waals materials, layered transition-metal dichalcogenides (TMDs) offer a rich setting in which magnetism, electronic correlations, and topology can coexist, hosting phenomena such as proposed spin-liquid behavior in the Mott insulator 1T-$TaS_2$ [11,12] and time-reversal-symmetry-breaking superconductivity in 4Hb-$TaS_2$ [13]. An intriguing member of this family is $Ni_{1/4}TaSe_2$, a layered correlated ferromagnet that lies close in composition to a superconducting phase [14] and whose crystal structure is shown in Fig. 1a. This proximity makes $Ni_{1/4}TaSe_2$ a compelling system in which to examine the relationship between magnetism, topology, and superconductivity.

$Ni_{1/4}TaSe_2$ exhibits ferromagnetic ordering below $T_C$ = 58 K [14]. Previous studies indicate that both Ni and Ta carry magnetic moments aligned along the *c* axis, as illustrated in Fig. 1b. The Ni moment is approximately 0.7 μB, whereas the Ta moments are considerably smaller, ranging from 0.05 to 0.12 μB depending on the crystallographic site [14]. The ferromagnetic transition is also evident in susceptibility and resistivity measurements [14], while the magnetic order is further supported by our x-ray magnetic circular dichroism (XMCD) measurements across the Ni $L_{2,3}$ edges (Fig. 1c). Remarkably, in Ni-depleted compositions, superconductivity emerges, with the resistivity dropping abruptly to zero below approximately 2 K [14].

The compositional proximity of the ferromagnetic and superconducting phases raises the question of whether features of the ferromagnetic electronic state are relevant to the superconducting pairing mechanism. Previous density functional theory (DFT) calculations predict strongly spin-polarized electronic states in ferromagnetic $Ni_{1/4}TaSe_2$ [14]. Our calculations further indicate finite Berry curvature associated with selected bands, motivating an experimental search for corresponding topological signatures. Whether these spin-polarized and Berry-curvature-related features are realized experimentally, however, remains unresolved.

Recent advances have established circular-dichroic and spin-resolved photoemission as powerful probes of Berry-curvature-related and spin-orbital phenomena in quantum materials. Theoretical work by Schüler *et al.* connected circular-dichroic ARPES to local Berry-curvature signatures and proposed its extension to spin-resolved measurements [15], while Di Sante *et al.* subsequently demonstrated spin-resolved spectroscopic sensitivity to finite spin Berry curvature in kagome metals [16]. A complementary circular-dichroic ARPES study has demonstrated the utility of crystal-mirror-plane geometries for probing local Berry-curvature signatures in two-dimensional materials [17]. Further, combined polarization-, spin-, and angle-resolved photoemission has enabled reconstruction of the quantum geometric tensor in CoSn [18]. Related polarization- and spin-sensitive photoemission approaches have revealed Berry-flux-associated spin and orbital textures in TaAs [19], spin-orbital chiral signatures in $Sr_2RuO_4$ [20], and an anomalous spin-optical helical response in $CsTi_3Bi_5$ [21]. Together, these studies demonstrate the broad sensitivity of polarization- and spin-resolved photoemission to quantum geometry, spin-orbital coupling, and broken symmetries in solids. Building on these developments, we apply spin-resolved circular-dichroic photoemission to the intrinsically ferromagnetic state of $Ni_{1/4}TaSe_2$. This extends spin-resolved circular-dichroic ARPES to an intrinsically ferromagnetic, globally time-reversal-symmetry-broken phase, enabling us to probe its Berry-curvature-related electronic structure.

Here, we combine spin- and angle-resolved photoemission spectroscopy (SARPES), linearly and circularly polarized synchrotron radiation, and DFT calculations to probe the spin-resolved electronic structure of $Ni_{1/4}TaSe_2$. We establish the presence of spin-polarized states crossing the Fermi level and experimentally reveal signatures of time-reversal-symmetry breaking in the ferromagnetic state. In addition, spin-resolved circular-dichroism measurements reveal signatures consistent with a finite spin Berry curvature and suggestive of local topological character [15,16].

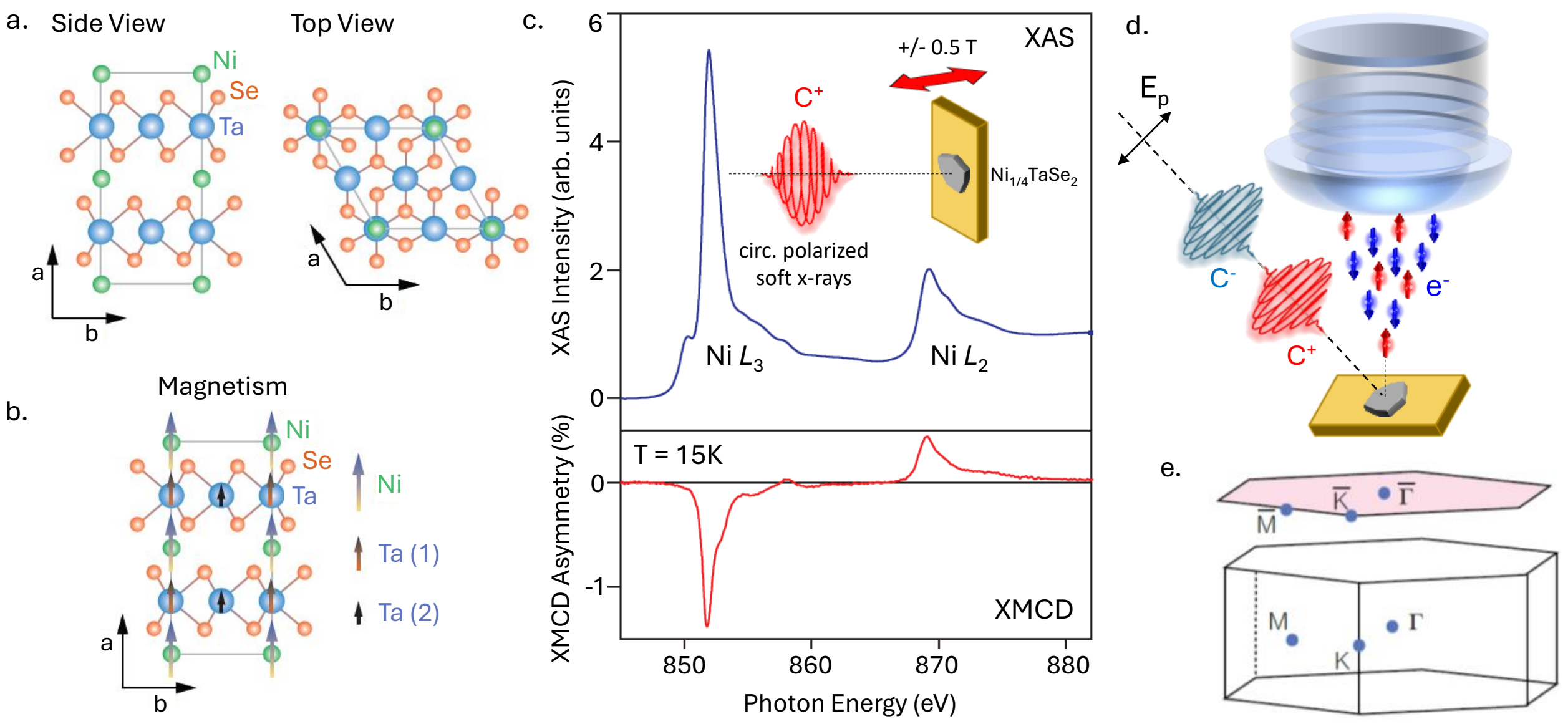

**Figure 1. Crystal and magnetic structures and experimental geometry of $Ni_{1/4}TaSe_2$. a,** Side and top views of the crystal structure. **b,** Ferromagnetic structure, with Ni and Ta moments aligned along the *c* axis; the Ni moment is approximately 0.7 μB, while the inequivalent Ta moments range from 0.05 to 0.12 μB. **c,** Ni $L_{2,3}$-edge x-ray absorption (XAS, top) and x-ray magnetic circular dichroism (XMCD, bottom), with the XMCD measurement geometry shown in the inset. **d,** SARPES geometry showing linearly and circularly polarized synchrotron radiation and spin-sensitive detection. **e,** Bulk Brillouin zone and surface projection, with the relevant high-symmetry points and directions indicated.

In this way, the spin-resolved dichroic response provides an experimental link between the ferromagnetic spin polarization and the Berry-curvature-related properties of the electronic states. Together, these results provide experimental support for the predicted spin-polarized electronic structure and Berry-curvature-related features of $Ni_{1/4}TaSe_2$, establishing a foundation for understanding its possible connection to superconductivity upon Ni depletion.

Single crystals of $Ni_{1/4}TaSe_2$ were synthesized using a two-step procedure described previously [14]. A polycrystalline precursor was first prepared from Ni, Ta, and Se, followed by iodine-assisted chemical vapor transport in a two-zone furnace, yielding millimeter-sized hexagonal single crystals. Spin- and angle-resolved photoemission spectroscopy (SARPES) measurements were performed at the APE Low-Energy beamline at Elettra using a VLEED detector coupled to a Scienta DA30 analyzer, with energy and momentum resolutions of approximately 12 meV and 0.018 Å$^{-1}$, respectively. Complementary momentum-microscopy measurements were performed at the NanoESCA laboratory at Elettra. The SARPES geometry, including the linear/circular polarization control and spin-sensitive detection, is shown schematically in Fig. 1d, while Fig. 1e shows the Brillouin zone and its surface projection used to define the measured high-symmetry directions. Measurements were performed at 15 K, except for the NanoESCA data collected at 95 K, with samples aligned along either $\bar{\Gamma}$–$\bar{K}$ or $\bar{\Gamma}$–$\bar{M}$. XAS and XMCD measurements were carried out at beamline 4.0.2 of the Advanced Light Source in total electron yield mode at 15 K, using circularly polarized x-rays in normal-incidence geometry and an applied magnetic field of ±0.5 T; the corresponding spectra and XMCD geometry are shown in Fig. 1c. DFT calculations were performed using VASP with projector-augmented-wave potentials and the SCAN meta-GGA functional, including spin-orbit coupling [22-24], while Wannier-based tight-binding models were constructed to calculate the Berry curvature [25].

We first established the underlying electronic structure of $Ni_{1/4}TaSe_2$ by comparing ARPES measurements with DFT calculations, providing a basis for testing the predicted spin-polarized electronic structure and Berry-curvature-related features of the ferromagnetic state. Figures 2a and 2b show the Fermi surface measured at 95 K, above $T_C$, and at 15 K, below $T_C$, respectively. The measured Fermi-surface contours are consistent with previous low-temperature ARPES measurements [14]. Within our data, the overall Fermi-surface structure changes little across the ferromagnetic transition. At both temperatures, a large, warped, nearly hexagonal pocket occupies much of the Brillouin zone, with the ARPES intensity exhibiting a threefold-symmetric modulation consistent with the underlying $C_3$ symmetry. Figure 2c shows the experimental energy-momentum dispersion along $\bar{K}'$–$\bar{\Gamma}$–$\bar{K}$. For comparison, Figs. 2d and 2e show the corresponding DFT dispersions along K′–Γ–K at $k_z$ = 0 in the nonmagnetic and ferromagnetic phases, respectively. The calculations reproduce the principal features of the measured dispersion and show the band splitting associated with the ferromagnetic state. The complementary nonmagnetic and ferromagnetic DFT dispersions along M′–Γ–M at $k_z = 0$ are shown in Fig. S1. Calculated Fermi surfaces in the ferromagnetic phase at $k_z = 0$, 0.25, and 0.5, including small shifts of the Fermi level used to facilitate comparison with experiment, are shown in Fig. S2.

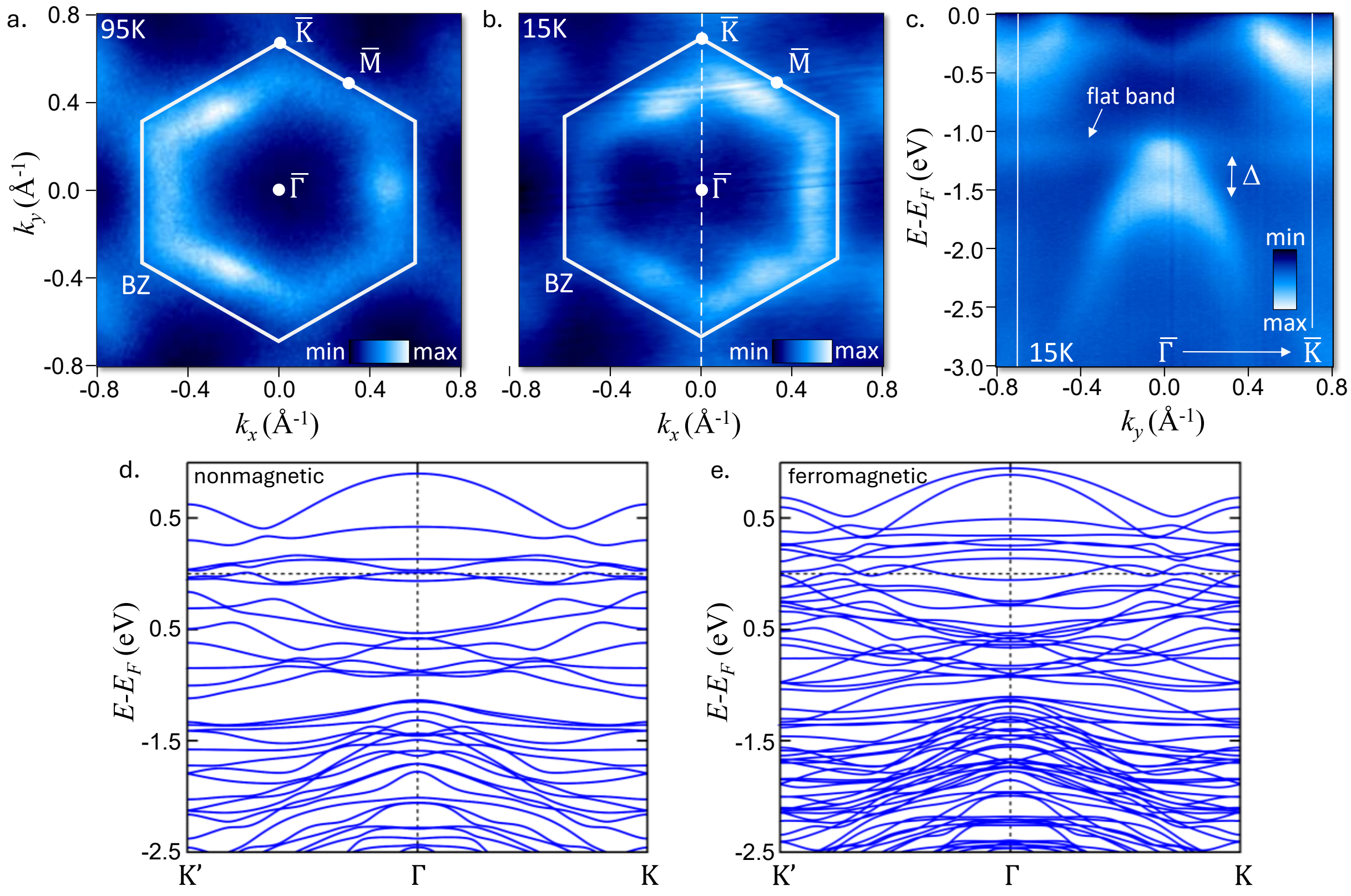

**Figure 2. Electronic structure of Ni1/4TaSe2 across the ferromagnetic transition. a,b,** ARPES Fermi surfaces measured with linearly polarized light at 95 K, above $T_C$ (a), and 15 K, below $T_C$ (b). **c,** ARPES energy-momentum dispersion along $\bar{K}'$–$\bar{\Gamma}$–$\bar{K}$ at 15 K. **d,e,** DFT band structures along K′–Γ–K at $k_z = 0$ in the nonmagnetic (d) and ferromagnetic (e) phases.

To determine the spin character associated with this measured band structure, we examined the ARPES spectra along the $\bar{\Gamma}$–$\bar{K}$ and $\bar{\Gamma}$–$\bar{M}$ directions in comparison with DFT. The $\bar{\Gamma}$–$\bar{K}$ cut is shown in Fig. 2c. The spectra reveal dispersive states crossing the Fermi level that are characteristic of metallic transition-metal dichalcogenides and broadly resemble those of the nonmagnetic parent compound $TaSe_2$. Prominent Fermi-level states are observed near both K̄ and M̄. DFT reproduces the principal low-energy features and predicts strong spin polarization in the ferromagnetic state, with predominantly majority-spin character near K̄ and minority-spin character near M̄. These states contain contributions from both Ta 4*d* and Ni 3*d* orbitals [14]. Figures 3a and 3b show the spin-resolved calculations at $k_z$ = 0 along K′–Γ–K and M′–Γ–M, respectively, revealing substantial spin polarization consistent with the *c*-axis magnetization.

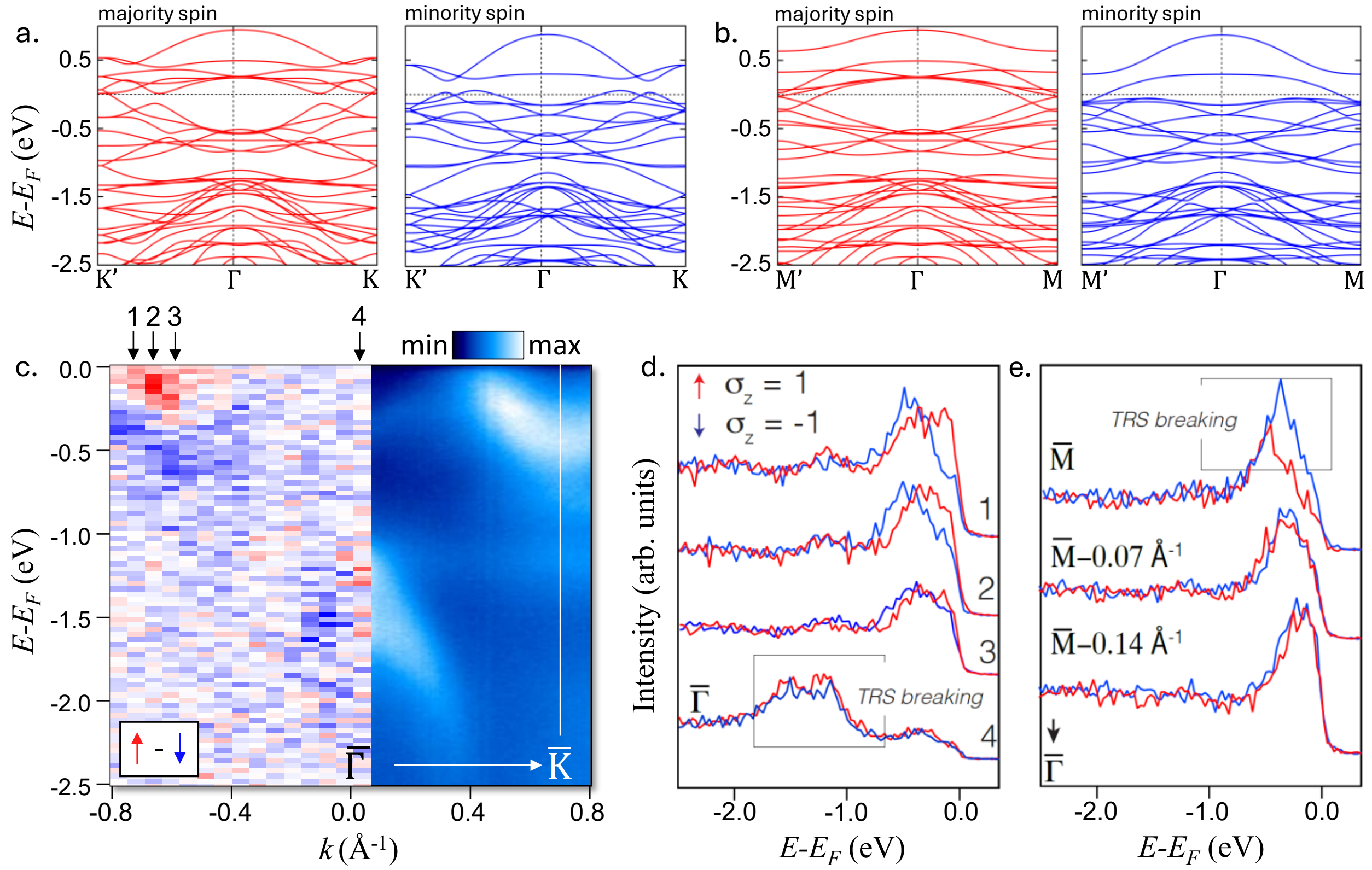

**Figure 3. Spin-polarized electronic structure of $Ni_{1/4}TaSe_2$. a,b,** Spin-resolved DFT band structures at $k_z$ = 0 along K′–Γ–K (a) and M′–Γ–M (b), showing the calculated spin-dependent electronic structure. **c,** Spin-polarization map and spin-integrated ARPES spectrum along Γ̄–K̄ measured with linear horizontal polarization. **d,e,** Representative $\sigma_z$-resolved energy distribution curves along Γ̄–K̄ (d) and Γ̄–M̄ (e). The finite out-of-plane spin polarization at the time-reversal-invariant Γ̄ and M̄ points is consistent with broken time-reversal symmetry in the ferromagnetic state.

The calculations also identify the Γ̄–K̄ direction as particularly favorable for experimentally testing the predicted spin polarization. As seen by comparing Figs. 3a and 3b, states of opposite spin character near the Fermi level are more clearly separated in energy and momentum along $\bar{\Gamma}$–$\bar{K}$ than along $\bar{\Gamma}$–$\bar{M}$, reducing their overlap within the experimental energy and momentum resolution. Because unresolved states with opposite spin polarization contribute simultaneously to the measured SARPES intensity, their signals partially cancel and reduce the observed polarization. We therefore focus primarily on the $\bar{\Gamma}$–$\bar{K}$ direction, while measurements along $\bar{\Gamma}$–$\bar{M}$ provide a complementary comparison.

Guided by these calculations, we used SARPES to probe the predicted spin polarization of the Fermi-level states. The incident beam was aligned within the crystal mirror plane $\bar{\Gamma}$–$\bar{M}$, and the out-of-plane spin component $\sigma_z$ was measured along momentum cuts in both the $\bar{\Gamma}$–$\bar{K}$ and $\bar{\Gamma}$–$\bar{M}$ directions. Linear horizontal polarization ($E_p$ in Fig. 1d), containing both in-plane and out-of-plane electric-field components in this geometry, was used to probe states with different orbital characters. Figure 3c shows the spin-polarization map alongside the spin-integrated photoemission spectrum along $\bar{\Gamma}$–$\bar{K}$ and indicates the energy-momentum region over which the out-of-plane spin polarization was measured.

To maximize sensitivity to the Fermi-level states, the SARPES measurements were performed at a photon energy of 62 eV, where favorable photoemission matrix elements enhance their spectral weight (Fig. S3a). Photon-energy-dependent ARPES measurements (Fig. S3a) show little discernible variation in the Fermi-level momentum structure over the measured photon-energy range, consistent with weak $k_z$ dispersion and the predominantly two-dimensional character of the observed states. An additional Fermi-surface map measured by momentum microscopy is shown in Fig. S3b.

Under these measurement conditions, the SARPES spectra reveal a pronounced out-of-plane spin polarization along $\bar{\Gamma}$–$\bar{K}$, together with a finite signal along $\bar{\Gamma}$–$\bar{M}$. Particularly significant is the observation of finite spin polarization at $\bar{\Gamma}$, a time-reversal-invariant momentum. In a nonmagnetic system with preserved time-reversal symmetry, Kramers theorem requires electronic states at this momentum to occur as degenerate time-reversed pairs. The observed spin polarization is therefore consistent with the lifting of this degeneracy in the ferromagnetic state. A finite spin polarization is likewise observed at $\bar{M}$, another time-reversal-invariant momentum, further supporting the presence of broken time-reversal symmetry. We therefore identify momentum-resolved spin-spectroscopic signatures in $Ni_{1/4}TaSe_2$ consistent with broken time-reversal symmetry in its electronic states, manifested as finite spin polarization at the time-reversal-invariant $\bar{\Gamma}$ and $\bar{M}$ points. These measurements provide spin-resolved experimental verification of the spin-polarized electronic structure previously predicted for ferromagnetic $Ni_{1/4}TaSe_2$ [14].

To examine the spectral origin of this spin signal in greater detail, Figs. 3d and 3e show representative spin-resolved energy distribution curves (EDCs) for $\sigma_z$ along the $\bar{\Gamma}$–$\bar{K}$ and $\bar{\Gamma}$–$\bar{M}$ directions, respectively. At $\bar{\Gamma}$, the spin signal retains the same sign across the two prominent spectral features. Although this might appear inconsistent with a simple picture of two oppositely spin-split bands, the calculated electronic structure contains several closely spaced bands within the same energy range. Because SARPES measures the combined spectral weight of states within its finite energy and momentum resolution, contributions from these additional bands modify the measured polarization. The resulting net signal can therefore retain the same sign across both spectral features, consistent with the calculated multiband electronic structure. Having established the spin character of these states, we next asked whether the ferromagnetic spin polarization is accompanied by a spin-dependent Berry-curvature response, which would provide a direct connection between magnetism and the Berry-curvature-related electronic structure suggested by theory.

To test this possibility, we performed circular-dichroic SARPES measurements following the theoretical framework developed by Schüler *et al.* [15] and its spin-resolved experimental implementation by Di Sante *et al.* [16]. To reduce contributions arising purely from the experimental geometry, the incident beam was confined to the Γ̄–M̄ mirror plane, while measurements were performed along Γ̄–K̄ [17]. Figure 4a shows the linearly polarized ARPES dispersion and identifies the region selected for these measurements. We focused on the split hole-like states near Γ̄ at approximately -1.5 eV, where symmetry considerations are expected to further reduce geometric contributions to the circular dichroism [16,17]. Resolving the dichroic response into the two spin channels can then provide sensitivity to the spin-dependent component associated with the local Berry curvature.

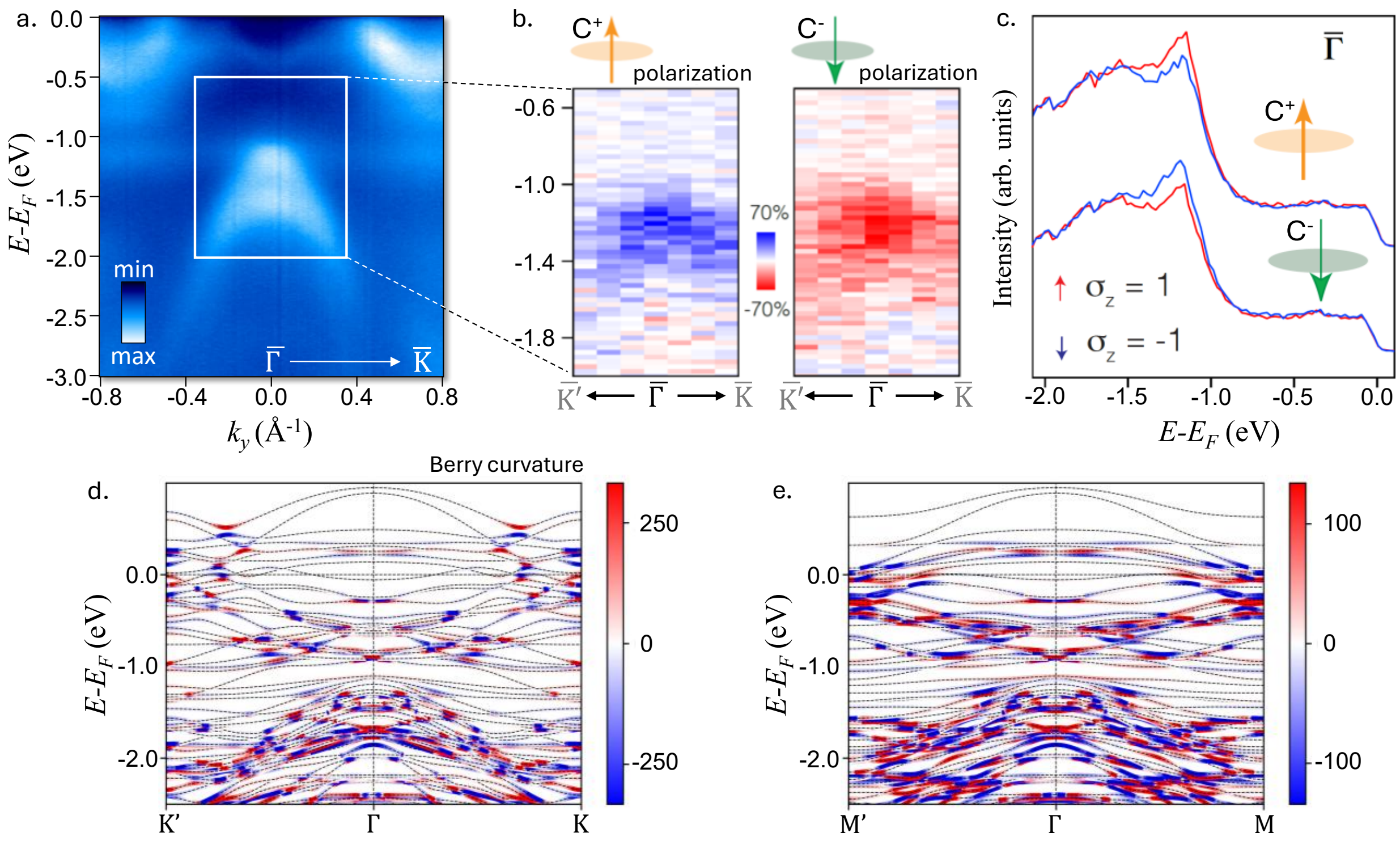

**Figure 4. Spin-resolved circular dichroism and local topological character in $Ni_{1/4}TaSe_2$. a,** Linearly polarized ARPES dispersion along K̄′–Γ̄–K̄, with the region containing the split hole-like states near Γ̄ highlighted for circular-dichroism measurements. **b,** Spin-polarization maps of the highlighted region measured with right- ($C^+$) and left-circularly polarized ($C^-$) light. **c,** Spin-resolved EDCs at Γ̄ measured with $C^+$ and $C^-$ light, showing a reversal of the spin-dependent response upon reversal of the photon helicity. **d,e,** Calculated ferromagnetic band structures with the Berry curvature projected onto the bands along K′–Γ–K (d) and M′–Γ–M (e), respectively.

Figure 4b shows the spin-polarization maps acquired with right- and left-circularly polarized light ($C^+$ and $C^-$). The corresponding spin- and helicity-resolved SARPES spectra underlying this analysis are shown in Fig. S4. A pronounced reversal of the spin-dependent response is observed upon reversing the photon helicity. This behavior is also evident in the spin-resolved energy distribution curves at $\bar{\Gamma}$ shown in Fig. 4c. For comparison, Figs. 4d and 4e show the calculated Berry-curvature-projected electronic structures along K′–Γ–K and M′–Γ–M, respectively, while complementary calculations near the Fermi level are shown in Fig. S5. The calculations reveal finite Berry curvature in the same energy range as the measured states. Although circular-dichroic photoemission can also be influenced by photoemission matrix elements and experimental geometry, the symmetry-controlled geometry used here is designed to minimize these contributions, particularly at $\bar{\Gamma}$, where the spin-resolved EDCs in Fig. 4c are evaluated. Within the framework of Di Sante *et al.* [16], the observed helicity-dependent spin response is therefore consistent with a finite spin Berry curvature and suggestive of local topological character, while not constituting a direct quantitative determination of the spin Berry curvature. Importantly, this response provides an experimental link between the spin-polarized ferromagnetic electronic structure and the local topological character associated with Berry curvature.

Taken together, our SARPES and circular-dichroism measurements, combined with DFT calculations, establish a coherent picture of $Ni_{1/4}TaSe_2$ as a layered ferromagnet with spin-polarized states crossing the Fermi level, signatures of broken time-reversal symmetry, and a helicity-dependent spin response consistent with a finite spin Berry curvature and suggestive of local topological character. The coexistence of these properties in a compositionally tunable system that becomes superconducting upon Ni depletion provides an opportunity to track how spin polarization and Berry-curvature-related signatures evolve toward the superconducting phase, potentially revealing connections between magnetism, topology, and superconductivity. More broadly, these results illustrate how spin polarization, magnetic order, and topological signatures can become intertwined in correlated van der Waals materials.

## Supporting Information

Additional calculated electronic structures and Fermi surfaces; photon-energy-dependent ARPES and momentum-microscopy data; spin- and helicity-resolved SARPES spectra; Berry-curvature calculations; and additional computational details (PDF).

**Notes:**

The authors declare no competing financial interest.

## ACKNOWLEDGMENTS

J.R.P., L.A.L., A.M.D., and A.X.G. acknowledge support from the U.S. Department of Energy, Office of Science, Office of Basic Energy Sciences, under Award No. DE-SC0024132. A.X.G. also gratefully acknowledges support from the Alexander von Humboldt Foundation. F.M. acknowledges MUR funding within the FIS2 (n. 1236, 1-8-2023) Project no. FIS-2023-03652 (CUP D53C25001950001). This work was partly performed within the framework of the Nanoscience Foundry and Fine Analysis (NFFA-MUR Italy, Progetti Internazionali) facility. The work at Northeastern University was supported by the National Science Foundation through the Expand-QISE award NSF-OMA-2329067 and benefited from the resources of Northeastern University's Advanced Scientific Computation Center, the Explorer Cluster, the Massachusetts Technology Collaborative award MTC-22032, and the Quantum Materials and Sensing Institute (QMSI). This research used resources of the Advanced Light Source, a U.S. Department of Energy Office of Science User Facility under Contract No. DE-AC02-05CH11231.

# Supporting Information for

# Hidden topological signatures in a van der Waals ferromagnet revealed by spin-resolved circular-dichroic photoemission

Jay R. Paudel[1,2], Federico Mazzola[3,4], Lidia A. Lapinski[1,5], Barun Ghosh[6,7], Valeria Rosa Rocha[8], Abigail M. Derrico[1,8], Nikola Maksimovic[8], Iulia Cojocariu[9], Daniel Baranowski[9], Christoph Klewe[10], Vitaliy Feyer[9], Ivana Vobornik[11], Claus M. Schneider[9], Giancarlo Panaccione[11], Arun Bansil[6,12], James G. Analytis[8], and Alexander X. Gray[1,*]

[1] *Department of Physics, Temple University, Philadelphia, Pennsylvania 19122, USA*

[2] *Chemical Sciences Division, Lawrence Berkeley National Laboratory, Berkeley, California 94720, USA*

[3] *Istituto di Struttura della Materia, Consiglio Nazionale delle Ricerche (CNR-ISM), 34149 Trieste, Italy*

[4] *Department of Physics and Astronomy "Galileo Galilei", University of Padova, 35131 Padova, Italy*

[5] *Department of Physics, Stanford University, Stanford, California 94305, USA*

[6] *Department of Physics, Northeastern University, Boston, Massachusetts 02115, USA*

[7] *Department of Condensed Matter and Materials Physics, S. N. Bose National Centre for Basic Sciences, Kolkata, India*

[8] *Department of Physics, University of California, Berkeley, California 94720, USA*

[9] *Peter Grünberg Institut (PGI-6), Forschungszentrum Jülich GmbH, 52428 Jülich, Germany*

[10] *Advanced Light Source, Lawrence Berkeley National Laboratory, Berkeley, California 94720, USA*

[11] *Istituto Officina dei Materiali, Consiglio Nazionale delle Ricerche (CNR-IOM), 34149 Trieste, Italy*

[12] *Quantum Materials and Sensing Institute, Northeastern University, Burlington, Massachusetts 01803, USA*

[*] *E-mail: axgray@temple.edu*

## Figure S1 | Effect of ferromagnetism on the calculated electronic structure along M′–Γ–M

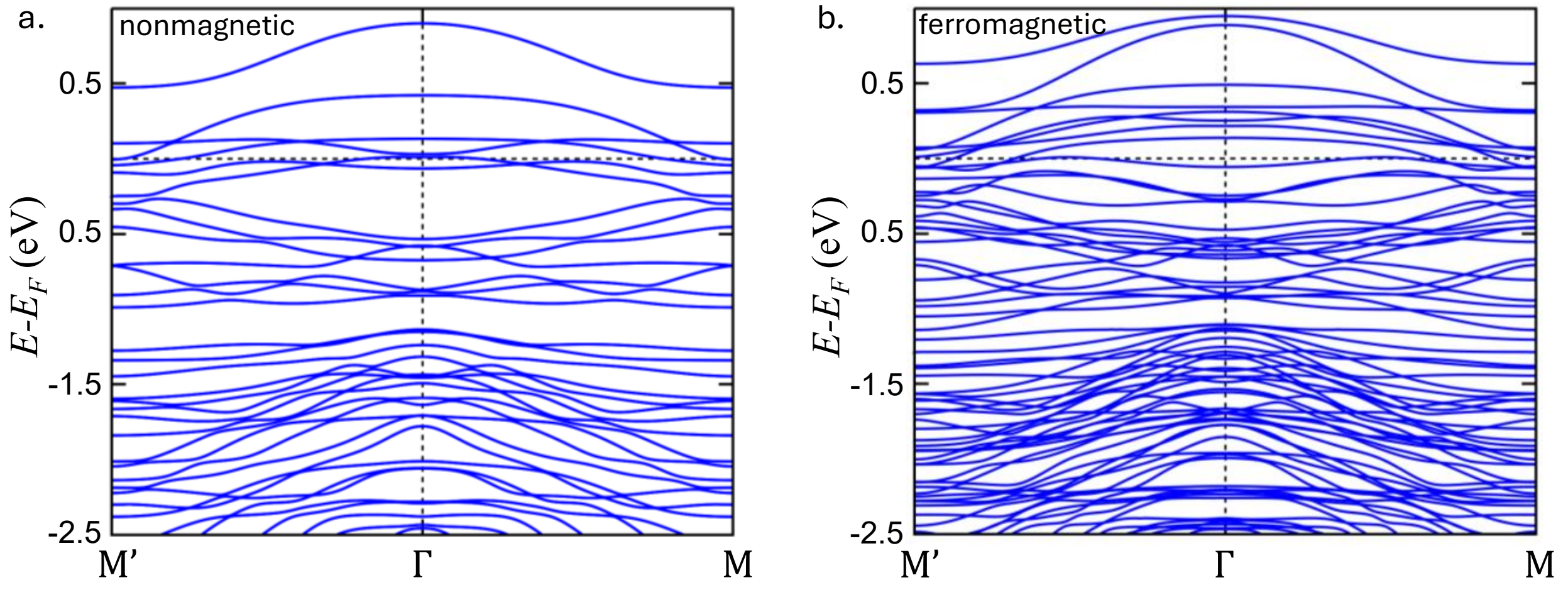

**Figure S1. a,b,** DFT band structures along M′–Γ–M at $k_z = 0$ in the nonmagnetic (a) and ferromagnetic (b) phases, respectively, complementing the corresponding K′–Γ–K calculations shown in Fig. 2d,e. We used a plane-wave kinetic-energy cutoff of 500 eV and a Γ-centered 10×10×6 *k*-point mesh.

## Figure S2 | Calculated Fermi surfaces of $Ni_{1/4}TaSe_2$

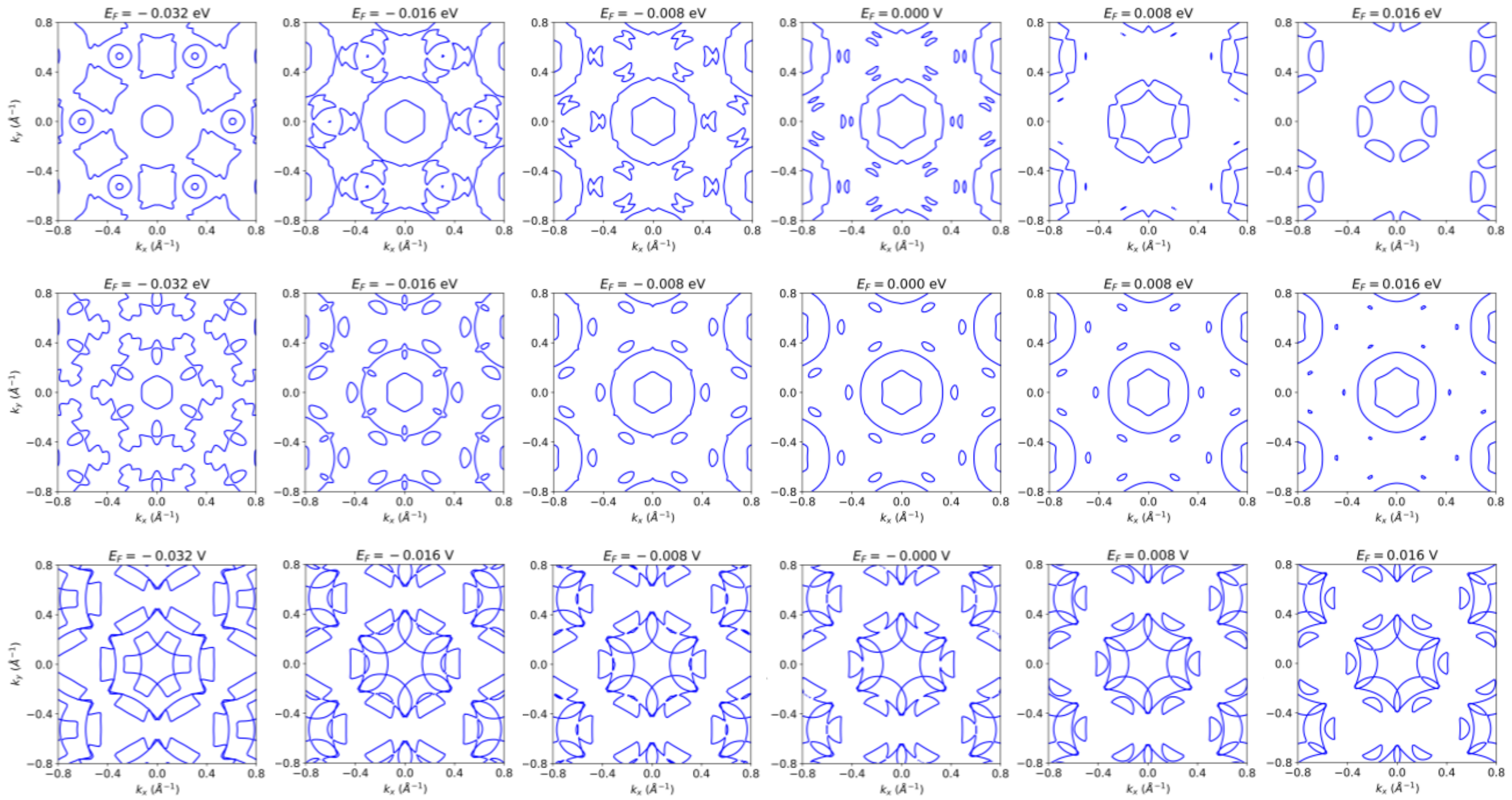

**Figure S2.** Calculated Fermi surfaces of $Ni_{1/4}TaSe_2$. Calculated Fermi-surface contours in the ferromagnetic phase at $k_z = 0$, 0.25, and 0.5 (top, middle, and bottom rows, respectively) for small shifts of the Fermi level relative to the calculated value, illustrating the dependence of the Fermi-surface structure on out-of-plane momentum and chemical potential. The same computational parameters as in Fig. S1 were used.

**Figure S3 | Photon-energy dependence and *k*-space electronic structure of $Ni_{1/4}TaSe_2$.**

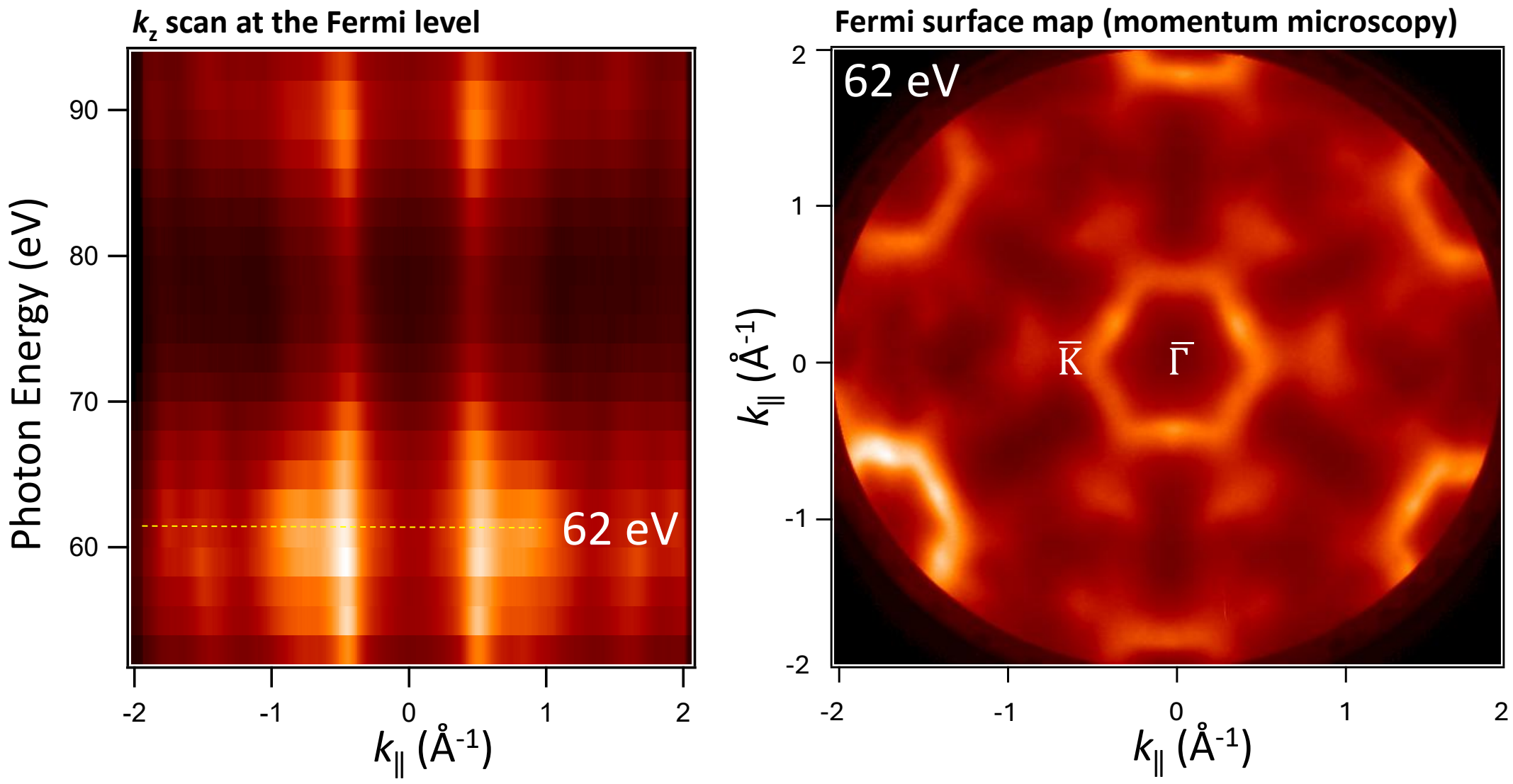

**Figure S3. a,** Photon-energy-dependent ARPES intensity at the Fermi level showing little variation in the momentum-space structure over the measured photon-energy range, consistent with weak $k_z$ dispersion and the predominantly two-dimensional character of the electronic states. **b,** Fermi-surface map measured by momentum microscopy at the NanoESCA laboratory at 95 K, above the ferromagnetic transition temperature.

**Figure S4 | Spin- and helicity-resolved photoemission spectra**

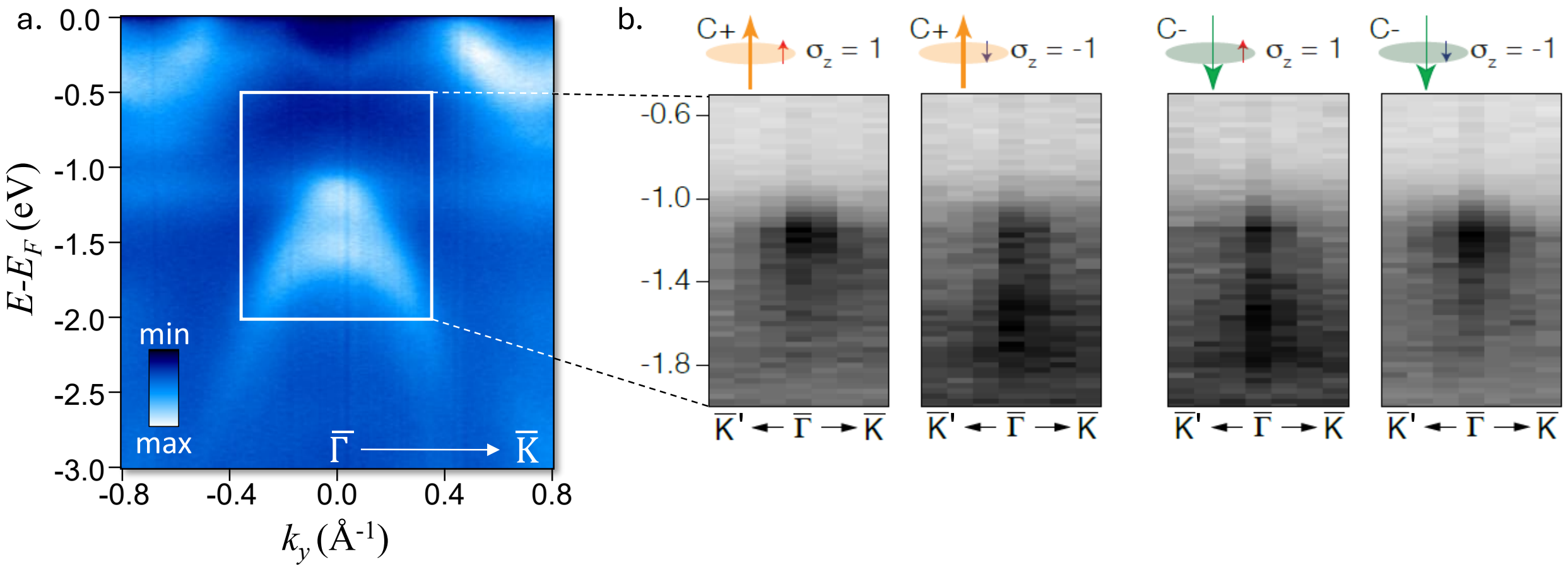

**Figure S4. a,** Linearly polarized ARPES dispersion along $\bar{K}'$–$\bar{\Gamma}$–$\bar{K}$, with the region containing the split hole-like states near $\bar{\Gamma}$ highlighted for the circular-dichroism measurements. **b,** Spin-resolved SARPES spectra of the highlighted region measured with right- ($C^+$) and left-circularly polarized ($C^-$) light for both $\sigma_z = +1$ and $\sigma_z = -1$ spin channels. These spectra provide the underlying spin- and helicity-resolved data used to obtain the spin-polarization maps and spin-resolved EDCs shown in Fig. 4b,c.

**Figure S5 | Berry-curvature-projected electronic structure of ferromagnetic $Ni_{1/4}TaSe_2$**

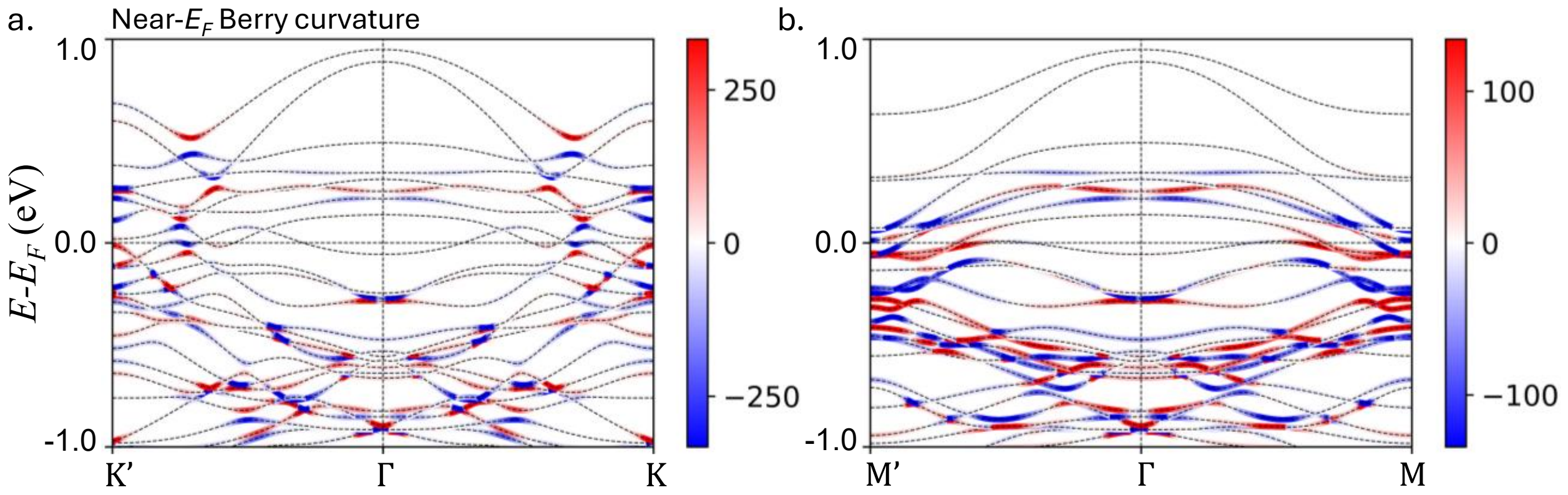

**Figure S5. a,b,** Calculated ferromagnetic band structures with the Berry curvature projected onto the bands along K′–Γ–K (a) and M′–Γ–M (b), shown over the near-Fermi-level energy range. Wannier-function-based tight-binding models were constructed using Ni *d*, Ta *s* and *d*, and Se *p* orbitals as the projection basis. The Berry curvature was calculated from these models using the standard Kubo formula as implemented in an in-house code. These calculations complement the broader-energy-range Berry-curvature projections shown in Fig. 4d,e.